# Hazard assessment of potential storm tide inundation at Southeast China coast

Bingchuan Nie [1,2], Qingyong Wuxi [3,4], Jiachun Li [3,4,*] and Feng Xu[1,2]

[1] School of Civil Engineering, Beijing Jiaotong University, Beijing 100044, China
[2] Beijing's Key Laboratory of Structural Wind Engineering and Urban Wind Environment, Beijing 100044, China
[3] Key Laboratory for Mechanics in Fluid Solid Coupling Systems, Institute of Mechanics, Chinese Academy of Sciences, Beijing 100190, China
[4] School of Engineering Science, University of Chinese Academy of Sciences, Beijing 100049, China

**Abstract** Hazard assessment of storm tide is addressed for Southeast China coast in this study. In particular, we pay attention to the scarcely discussed issue of storm tide inundation. The main procedures of hazard assessment are: 1) non-stationary tropical cyclone intensification (TCI) and sea level rise (SLR) for the study area are analyzed based on the long-term historical database; 2) four typical scenarios of storm tide are examined using the surge-tide-wave coupled hydrodynamic model; 3) the potential inundated regions are identified on the GIS platform. The distributions of water elevation show that high water elevation tends to occur in the bays and around the estuaries. Without considering the impacts of TCI and SLR, the maximal water elevations caused by the typhoon wind of 100-year recurrence period can reach as high as 6.06 m, 5.82 m and 5.67 m around Aojiang, Feiyunjiang and Oujiang river estuaries, respectively. In this circumstance, about 533 km$^2$ area is under the threat of storm tide inundation. TCI and SLR due to climate change can further deteriorate the situation and enhance the risk of inundation there. The maximal water elevations at the aforementioned three estuaries considering the potential non-stationary TCI and SLR in 2100s can be as high as 7.02m, 6.67 m, 6.44m, respectively. The corresponding potential inundation area expands by 50% compared with the situation without considering TCI and SLR. In addition, the remotely sensed maps and inundation durations of the hardest hit regions are provided as well.

**Keywords** storm surge; inundation; risk assessment; tropical cyclone intensification; sea level rise

*E-mail: jcli08@imech.ac.cn

# 1. Introduction

Storm surge caused by tropical cyclone (TC) is one of the most hazardous events for coastal zones. It has produced devastating damage in low-lying areas worldwide in history (Li and Nie, 2017). Although the operational forecast of storm surge has made a great progress in the past decades, a few individual disastrous storm events still caused extensive damages recently, say, TC Haiyan in 2013 resulted in 6300 dead and 1061 missing. With frequent invasion of TC at the coastal area, China is heavily subjected to storm tide hazard. The annual average direct economic loss caused by storm tide is about 1.73 billion USD, and no downtrend can be observed over the last three decades. With more population and production activities in coastal areas owing to urbanization and industrialization, probably, the potential losses by storm tide in China can be more dramatic in the future.

Inundation is one of the most catastrophic consequences in a storm tide event. To reduce the potential losses during inundation, risk assessment is an effective and practical way. Based on the empirical model, Hsu et al. (2018) examined the risk of northern Gulf of Mexico coast exposed to storm surge. Christie et al. (2018) investigated the coastal flood risk in North Norfolk, in which CS3X Continental Shelf tidal surge model and 1D SWAN model are adopted. Younus (2017) analyzed the vulnerability issues during storm surge, fifty-six vulnerability issues are identified, and their categories and weighted index scale are provided. Based on a surrogate model, Taflanidis et al. (2013) carried out the risk estimation of TC waves, water elevations, and run-up for TC passing by the Island of Oahu. In those works, risk assessment is usually divided into hazard and vulnerability assessments. The vulnerability assessment is devoted to figure out the resistant capability of coastal zones towards storm surge inundation. While, the hazard assessment is aimed to evaluate the natural attributes of storm surge inundation, which is the prerequisite of vulnerability assessment.

For more accurate hazard assessment, the precision of storm tide prediction plays a pivotal role. Many researchers were devoted to developing hydrodynamic models of high performance for storm tide prediction. Jelesanski et al. (1992) built the popular SLOSH model which is still extensively used in operational forecast of storm surge. The continual renovation ADCIRC model developed by Luettich and his colleagues (1992) has been widely applied in the academic communities. Other hydrodynamic models such as POM (Peng et al. 2004), ROMS (Li et al. 2006), ELCIRC (Stamey qt al., 2007), FVCOM (Weisberg et al., 2008), SELFE (Shen et al., 2009) and CH3D (Sheng et al., 2010) have been used for storm surge studies as well. The aforementioned hydrodynamic storm surge models have been further upgraded to couple with the wave models recently. The typical works are: the ADCIRC+SWAN model by Dietrich et al. (2011; 2012), the FVCOM-SWAVE model by Wu et al. (2011)

and the SELFE-WWM-II by Roland et al. (2012). Based on those coupled models, the surge-tide-wave coupling effects can be readily presented. Zhang et al. (2017) concluded that the deviation of water elevation of order of one meter can be caused owing to nonlinear tide-surge interaction. Wuxi et al. (2018) observed wave induced surge and tide-surge interaction along Zhe-Min coast. They found that the relative error due to tide-surge coupling effect can be as high as 28% maximal. These facts imply that the surge-tide-wave coupled models are preferred in hazard assessment of storm tide inundation.

In addition to the precision of hydrodynamic model, the challenging issues due to climate change need to be considered for long-term hazard assessment. Tropical cyclone intensification (TCI) and sea level rise (SLR) are obvious among the direct influential factors. Recently, the impacts of TCI and SLR on storm surge in a few regions have been investigated. Feng and Li et al. (2018) examined the storm surge trends in the coastal areas of China from 1997 to 2016. They concluded that the increasing rate of extreme storm surge is as high as 0.06 m per year at 90% confidence level. Li et al. (2018) studied the coastal flood hazards at Oahu for 24 TCs in 2080–2099 by the CMIP5 NCAR-CCSM4 model and the SLR under Representative Concentration Pathway (RCP) 8.5. Wang et al. (2012) simulated the inundation in Shanghai considering SLR, land subsidence. They found that 46% of the seawalls and levees are projected to be overtopped by 2100. Feng and Gao et al. (2018) investigated the inundation risk of extreme water elevations in Rongcheng based on the Pearson Type III distribution considering SLR under situations RCPs 2.6, 4.5 and 8.5. Yin et al. (2017) studied the impacts of SLR and TCI artificially designed on storm surges and waves at Pearl River Estuary, and demonstrated that effects of SLR and TCI act non-linear and spatially non-uniform.

Based on the aforementioned facts, a procedure for long-term hazard assessment of storm tide inundation is established relied on surge-tide-wave coupled model, local TCI and SLR database and geographic information system (GIS). Hazard assessment of storm tide inundation is implemented for Southeast China coast, one of the storm surge prone areas in China where TCs make landfall tending to have stronger intensity. The literature review shows that inundation caused by storm tide is scarcely studied, let alone the long-term hazard assessment of storm tide inundation considering potential TCI and SLR in the future at Southeast China coast. In the following sections, the surge-tide-wave coupled model is described in Section 2; the local TCI and SLR at the study area are addressed in Section 3; the flow chart for long-term hazard assessment of storm tide inundation are formulated in Section 4; and the hazard assessment results are presented and discussed in Section 5. Finally, Section 6 brings this study to an end with a few conclusions.

## 2. Hydrodynamic surge-tide-wave coupled model

### 2.1 Governing equations

The water elevation $\zeta$ of storm tide is described by the Generalized Wave Continuity Equation (GWCE) as (1). The vertical uniform horizontal current velocity $U$ and $V$ are governed by the vertically-integrated momentum equations, i.e. (2) and (3).

$$\frac{\partial^2 \zeta}{\partial t^2} + \tau_0 \frac{\partial \zeta}{\partial t} + S_p \frac{\partial J_\lambda}{\partial \lambda} + \frac{\partial J_\varphi}{\partial \varphi} - S_p UH \frac{\partial \tau_0}{\partial \lambda} - VH \frac{\partial \tau_0}{\partial \varphi} = 0, \tag{1}$$

$$\frac{\partial U}{\partial t} + S_p U \frac{\partial U}{\partial \lambda} + V \frac{\partial U}{\partial \varphi} + \frac{\partial J_\varphi}{\partial \varphi} - fV = -gS_p \frac{\partial}{\partial \lambda}\left[\zeta + \frac{P_s}{g\rho_0} - \alpha\eta\right] \\ + \frac{\tau_{s\lambda,winds} + \tau_{s\lambda,waves} - \tau_{b\lambda}}{\rho_0 H} + \frac{M_\lambda - D_\lambda}{H}, \tag{2}$$

$$\frac{\partial V}{\partial t} + S_p U \frac{\partial V}{\partial \lambda} + V \frac{\partial V}{\partial \varphi} + \frac{\partial J_\varphi}{\partial \varphi} - fU = -gS_p \frac{\partial}{\partial \varphi}\left[\zeta + \frac{P_s}{g\rho_0} - \alpha\eta\right] \\ + \frac{\tau_{s\varphi,winds} + \tau_{s\varphi,waves} - \tau_{b\varphi}}{\rho_0 H} + \frac{M_\varphi - D_\varphi}{H}. \tag{3}$$

Where $\lambda$ and $\varphi$ are the longitude and latitude, respectively. $H$ is the sum of water elevation and bathymetric depth. $M_\lambda$ and $M_\varphi$ are the vertically-integrated lateral stress gradients. $D_\lambda$ and $D_\varphi$ are the momentum dispersion terms. $S_P$ is a spherical coordinate conversion factor. $\tau_{s\lambda,winds}$ (or $\tau_{s\varphi,winds}$) and $\tau_{b\lambda}$ (or $\tau_{b\varphi}$) are the winds and bottom friction stresses, respectively.

$J_\lambda$ and $J_\varphi$ in (1) to (3) can be rewritten as (4) and (5), respectively (Dietrich et al. 2011).

$$J_\lambda = -S_p Q_\lambda \frac{\partial U}{\partial \lambda} - Q_\varphi \frac{\partial U}{\partial \varphi} + fQ_\lambda - \frac{g}{2} S_p \frac{\partial \zeta^2}{\partial \lambda} - gS_p H \frac{\partial}{\partial \lambda}[\frac{P_s}{g\rho_0} - \alpha\eta] \\ + \frac{\tau_{s\lambda,winds} + \tau_{s\lambda,waves} - \tau_{b\lambda}}{\rho_0} + (M_\lambda - D_\lambda) + U \frac{\partial \zeta}{\partial t} + \tau_0 Q_\lambda - gS_p H \frac{\partial \zeta}{\partial \lambda}, \tag{4}$$

$$J_\varphi = -S_p Q_\lambda \frac{\partial V}{\partial \lambda} - Q_\varphi \frac{\partial V}{\partial \varphi} - fQ_\lambda - \frac{g}{2}\frac{\partial \zeta^2}{\partial \varphi} - gH\frac{\partial}{\partial \varphi}[\frac{P_s}{g\rho_0} - \alpha\eta]$$
$$+ \frac{\tau_{s\varphi,winds} + \tau_{s\varphi,waves} - \tau_{b\varphi}}{\rho_0} + (M_\varphi - D_\varphi) + V\frac{\partial \zeta}{\partial t} + \tau_0 Q_\varphi - gH\frac{\partial \zeta}{\partial \varphi}. \tag{5}$$

Here, $g$, $\alpha$, $\rho_0$ and $\eta$ are the gravitational acceleration, the tidal potential, water density and the effective earth elasticity factor, respectively. $Q_\lambda$ (or $Q_\varphi$) is the product of current velocity and $H$.

The wave action density $N$ is conserved during propagation in the presence of ambient current. Hence, the wind wave satisfies the balance equation of the wave action density given as (6). $c_\lambda$ (or $c_\varphi$) is the group velocity of water wave. $c_\sigma$ and $c_\theta$ are the propagation speeds in spectral space. The source term $S_{tot}$ is usually constituted of six processes which can be as written as (7) (Zijema, 2010). $S_{in}$ is the wave energy inputted by the wind. $S_{ds,w}$, $S_{ds,b}$ and $S_{ds,br}$ are wave dissipation terms caused by white capping, bottom friction and depth-induced wave breaking, respectively. $S_{nl,3}$ and $S_{nl,4}$ represent the three-wave and four-wave interactions, respectively.

$$\frac{\partial N}{\partial t} + \frac{\partial}{\partial \lambda}[(c_\lambda + U)N] + \cos^{-1}\varphi\frac{\partial}{\partial \varphi}[(c_\varphi + V)N\cos\varphi]$$
$$+ \frac{\partial}{\partial \theta}[c_\theta N] + \frac{\partial}{\partial \sigma}[c_\sigma N] = \frac{S_{tot}}{\sigma}. \tag{6}$$

$$S_{tot} = S_{in} + S_{ds,w} + S_{ds,b} + S_{ds,br} + S_{nl3} + S_{nl4}. \tag{7}$$

The parameterization of wave energy input/dissipation terms, i.e. $S_{in}$, $S_{ds,w}$, $S_{ds,b}$ and $S_{ds,br}$, are expressed as (8) to (11). Where, $A_{in}$ and $B_{in}$ satisfy the linear growth formula (see Cavaleri et al., 1981) and exponential growth formula (see Komen et al., 1984), respectively. $\Gamma$ is a coefficient dependent on water depth. $C_b$ is the bottom friction coefficient. $E_{tot}$ and $D_{tot}$ are the total wave energy and the rate of dissipation due to wave breaking, respectively.

$$S_{in} = A_{in} + B_{in}E(\sigma,\theta), \tag{8}$$

$$S_{ds,w} = -\Gamma\overline{\sigma}\frac{k}{\overline{k}}E(\sigma,\theta), \tag{9}$$

$$S_{ds,b} = -C_b \frac{\sigma^2}{g^2 \sinh^2 kH} E(\sigma,\theta), \tag{10}$$

$$S_{ds,br} = \frac{D_{tot}}{E_{tot}} E(\sigma,\theta). \tag{11}$$

The coupling manner between the wave and water elevation is that: wind wave contributes to the water elevation via wave radiation stresses, i.e. $\tau_{s\lambda,waves}$ and $\tau_{s\varphi,waves}$ in (2)-(5); while the water elevation affects the wave dissipation terms (8)-(11) in turn.

In this study, the ADCIRC+SWAN model developed by Dietrich et al. [2011; 2012] is used to solve the aforementioned coupled governing equations. The continuous-Galerkin, finite-element model ADCIRC solves (1)-(5) to obtain the time dependent water elevation and current (Jelesnianski et al., 1992). The third-generation wave model SWAN estimates the evolution of wave parameters by (6)-(11) (Boij et al., 1999). In every interval, ADCIRC accesses the gradients of radiation stresses by S7WAN, while SWAN accesses the current and water elevation provided by ADCIRC.

## 2.2 External forcing and dissipation

The hydrodynamic surge-tide-wave model described by (1) to (11) is driven by the stresses due to wind $\tau_{s\lambda,winds}$ ($\tau_{s\varphi,winds}$) and bottom friction $\tau_{b\lambda}$ ($\tau_{b\varphi}$), the pressure $P_s$, the water elevation and current of astronomic tide. They are described as follows.

The wind stress $\tau_{s\lambda,winds}$ ($\tau_{s\varphi,winds}$) is determined by the drag coefficient $C_d$ and the wind field $V_{wind}$ at 10 m above sea surface. Here, $C_d$ increasing linearly with wind speed as proposed by Garrat (1977) is used. The wind field $V_{wind}$ is reconstructed based on the Holland model (Hubbert et al., 1991; Jakobsen et al., 2004), which can be decomposed into tangential wind velocity $V_T$, radial wind velocity $V_r$ and environmental scale wind velocity $V_E$, see (12). The expressions of $V_T$, $V_r$ and $V_E$ are given by (13)-(15), respectively.

$$V_{wind} = V_T + V_r + V_E. \tag{12}$$

$$V_T = \frac{1}{V_m}(\sqrt{R^{-B}\exp(1-R^{-B}) + a^2 R^2} - aR). \tag{13}$$

$$V_r = \frac{B[BR^{-2B} + (1-3B)R^{-B} + B - 1]k - 2k - RC}{B(R^{-B} - 1) + 2 + 4aR(V_T/V_m)^{-1}} V_T. \tag{14}$$

$$V_E = U_0 \exp\left(-\frac{r}{R_G}\right). \tag{15}$$

Here, $R$ equals to $r/R_M$, $r$ the distance to the TC eye. $V_m$, $U_0$, $R_G$ and $R_M$ are the maximal wind velocity, the migration speed, the length scale and the radius of maximum wind of TC, respectively. $B$, $C$ and $k$ are coefficients, which equals to 1.5, 0.013 and 0.16, respectively. $a$ demotes $fR_M/(2V_m)$, where $f$ the Coriolis parameter.

The pressure field is calculated as below (Hubbert et al., 1991; Jakobsen et al., 2004)

$$P = P_c + (P_n - P_c)\exp(-R^{-B}). \tag{16}$$

Once the maximal wind speed $V_m$, the central pressure $P_c$, the ambient pressure $P_n$ and the track of TC are given, the wind stresses $\tau_{s\lambda,winds}$ and $\tau_{s\varphi,winds}$ and the pressure $P_s$ can be reconstructed based on (12)-(16).

The bottom friction stress $\tau_{b\lambda}$ ($\tau_{b\varphi}$) equals the product of density $\rho_0$, $K_{slip}$ and the square of current $U^2$ ($V^2$). $K_{slip}$ is the bottom friction coefficient given by

$$K_{slip} = C_{f\min}[1 + (\frac{H_{break}}{H})^{\theta_f}]^{\gamma_f/\theta_f} (U+V)^{1/2}. \tag{17}$$

Where, $H_{break}$ is the wave break depth. $C_{f\min}$, $\theta_f$ and $\gamma_f$ are set to the recommended values, i.e. 0.0026, 10 and 1/3, respectively (Luettich et al., 2006).

Eight constituents K1, K2, M2, N2, O1, P1, Q1 and S2 from the Le Provost tidal database FES95.2 are used. The tidal elevation and depth-averaged tidal current are exerted on the open boundary.

The aforementioned parameterization has been validated for TCs Thane on the India Ocean and Saomai on the Northwest Pacific Ocean. Water elevation, significant wave height and surge are compared with in-situ data, and good agreements have been observed, for which one may refer to Wuxi et al. (2018).

## 3. Studied area

### 3.1 Computational domain

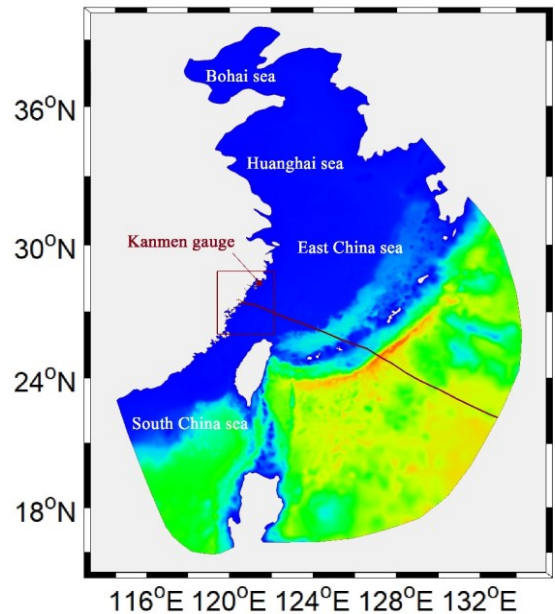

**Figure 1.** Overview of the computational domain. The region by the red rectangle is the area of interest. The coordinates of the corners of the red rectangle are (119.224°E, 28.687°N), (122.049°E, 28.687°N), (119.224°E, 25.545°N), (122.049°E, 25.545°N), respectively. The red dot (28.083 °N, 121.128 °E) is the location of Kanmen tidal gauge.

Southeast China coast as indicated by the red rectangle in figure 1 is the area of interest. The whole computation domain is extended to the Northwest Pacific Ocean (NWP) and South China Sea (SCS) to exclude the impact of the open boundary. The GEBCO bathymetry database of 30 arc second resolution is adopted. Unstructured grids with varying density, about 105k elements, are adopted for the computational domain. The mesh scale at the area of interest is less than 100 m.

### 3.2 TCI and SLR

TCI and SLR are the most direct influential factors for long-term hazard assessment of storm tide inundation.

As sea surface temperature (SST) increases, the ocean will have more energy to convert to TC wind. Webster et al. (2005) examined the intensity of TC. They pointed out that TCs in the strongest Saffir-Simpson categories 4 and 5 have almost doubled in both number and proportion in the period 1970–2004 for all ocean basins. Relied on the satellite-based estimation of TC intensity, Elsner et al. (2008) reported that significant wind speed upward trends of 0.3±0.09 ms$^{-1}$ yr$^{-1}$ can be observed for the strongest TCs. The study by Knutson et al. (2010) shows that the globally averaged intensity of TC could increase 2–11% by 2100 due to greenhouse warming. Recently, Wang and Li (2016) examined the extreme wind speeds in SCS and NWP, and announced that the spatial inhomogeneous and non-stationary effects should be considered.

Relied on the non-stationary model (Wang and Li, 2016), the TC intensities of different recurrence periods over the study area are investigated. The basic conception of the non-stationary model is that the statistical parameters of TC wind vary with time. The main steps for estimating the extreme wind speed of TC are described as below: 1) the 72 years (1945–2013) TC database from JTWC is divided into 23 periods with 50 years in each period, i.e. 1945-1994, 1946-1995, …, 1967-2013; 2) the shape and scale parameters of Weibull distribution for the wind speed of TC affect the area of interest are estimated for each period; 3) the time series of statistical parameters are fitted; 4) finally, the extreme wind speeds of different return periods can be obtained based on the extreme value theory as presented in figure 2. Taking the maximal wind speed of 50-year and 100-year recurrence period as examples, they can be as high as 75.49 m/s and 78.48 m/s, respectively.

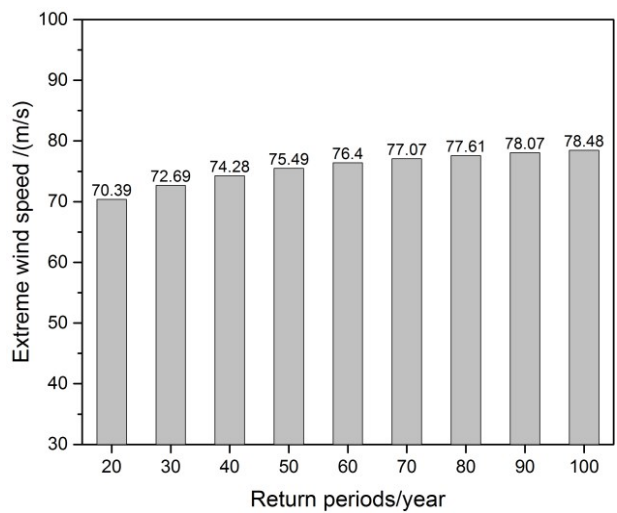

**Figure 2.** The extreme wind speed of TC with different recurrence periods for the study area considering non-stationary effect.

The AR5 of IPCC reported that the global SLR for 2081–2100 relative to 1986–2005 will likely be 0.26 to 0.55 m for RCP 2.6, 0.32 to 0.63 m for RCP 4.5, 0.33 to 0.63 m for RCP 6.0, and 0.45 to 0.82 m for RCP 8.5 (IPCC, 2015). They also claimed about 70% of the coastlines worldwide would experience lower SLR than the global mean SLR by 2100s. Considering the significant regional differences of SLR, the observed sea level series of Kanmen tidal gauge (28.083 °N, 121.128 °E), located in the study area (see figure 1), from the Permanent Service for Mean Sea Level

(PSMSL, www.psmsl.org) are analyzed. The monthly, one-year and five-year averaged MSLs from 1959 to 2017 are presented in figure 3. Obvious increasing trend of SLR can be observed. Based on the in-situ data, polynomial fitting is carried out as indicated by the red solid line in figure 3. The extrapolated potential SLR for 2050s is 0.185 m relative to 2006. While that for 2100s is 0.514 m, which is between the situations of RCP 6.0 and RCP 8.5.

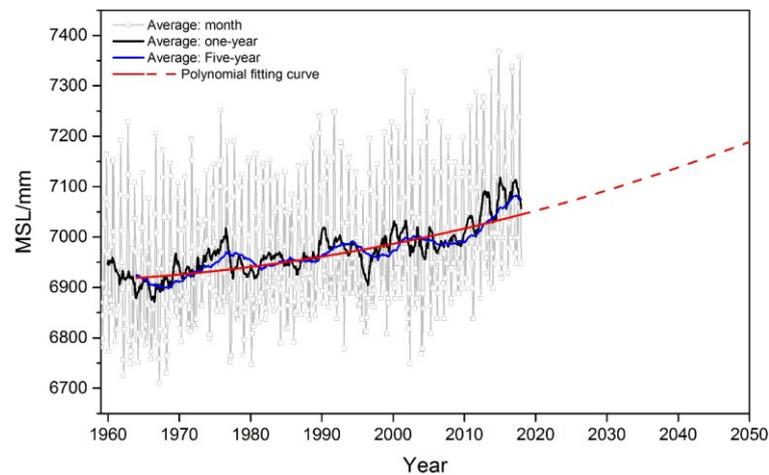

**Figure 3.** MSL series of Kanmen tidal gauge. The gray, black and blue solid lines are the averaged sea levels over one month, one year and five years, respectively. The red solid line is polynomial fitting result of the five-year averaged MSL, while the red-dash line is extrapolation of the fitting result.

## 4. Procedures of inundation hazard assessment

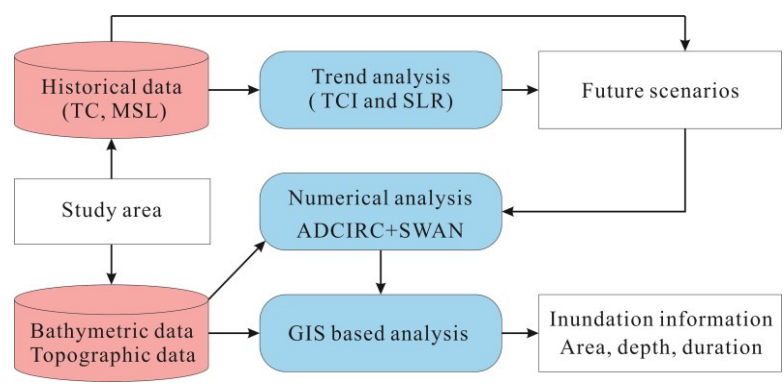

**Figure 4.** Flow chart for hazard assessment of storm tide inundation. The red cylinders and blue rounded rectangles represent the data sets and analytical approaches, respectively.

The procedures for hazard assessment of storm tide inundation under TCI and SLR are described by the flow chart in figure 4. The main steps are:

(1) Based on the history data, the non-stationary TCI and SLR can be obtained.

The parameters of the scenarios concerned can be presented.

(2) Storm tide of the scenarios concerned are simulated using the surge-tide-wave model described in Section 2. Distribution and evolution of water elevation can be obtained.

(3) Water elevation is then analyzed on the GIS platform, low-lying regions of inundation risk can be identified, and inundation hazard information including the area, inundation depth and duration for those regions can be figured out.

Saomai is the strongest TC made landfall at the study area. It struck the coast almost vertically as sketched by the solid line in figure 1. According to the National Meteorological Centre of China, the maximum wind speed of Saomai is about 60.0 m/s of 100-year recurrence period during landing. Therefore, scenario S2 (see table 1) is used to examine the storm tide inundation hazard without considering TCI and SLR. To evaluate the worst situation, typhoon makes landfall during the astronomical high tide in S2. Meanwhile, scenario S1, the hindcasting of actual situation of storm tide by Saomai, is also investigated.

Owing to the significant regional differences of TCI and SLR, the TCI and SLR based on analyzing the historical database of the study area are used in this work instead of using the global mean SLR under different RCPs and TCI artificial gave in the past studies. S3 and S4 as given in table 1 are investigated, which correspond to the situations 2050s and 2100s considering non-stationary TCI and SLR, respectively. According to the statistical results presented in Section 3.2, SLR of 0.185 m (0.514 m) compared with 2006 is adopted for S3 (S4). As for TCI, the extreme wind speeds of 50-year recurrence period and 100-year recurrence period are adopted for S3 and S4, respectively.

**Table 1**. Parameters of the scenarios concerned. Scenario S1 is the actual situation of TC Saomai. Scenario S2 is the situation that typhoon of wind speed of 100-year recurrence period makes landfall during the astronomical high tide without considering TCI and SLR. S3 (S4) corresponds to the situation that typhoon of wind speed of 50-year (100-year) recurrence period makes landfall during the astronomical high tide taking into account of the non-stationary TCI and SLR.

| Scenarios | MSL | TC intensity | Landing moment | TC track |
|---|---|---|---|---|
| S1 | 7013.6 mm | 60.0 m/s | Landing moment of Saomai | Saomai |
| S2 | 7013.6 mm | 60.0 m/s | High tide | Saomai |
| S3 | 7198.7 mm | 75.5 m/s | High tide | Saomai |

| S4 | 7526.6 mm | 78.5 m/s | High tide | Saomai |

## 5. Results and discussion

### 5.1 Water elevation

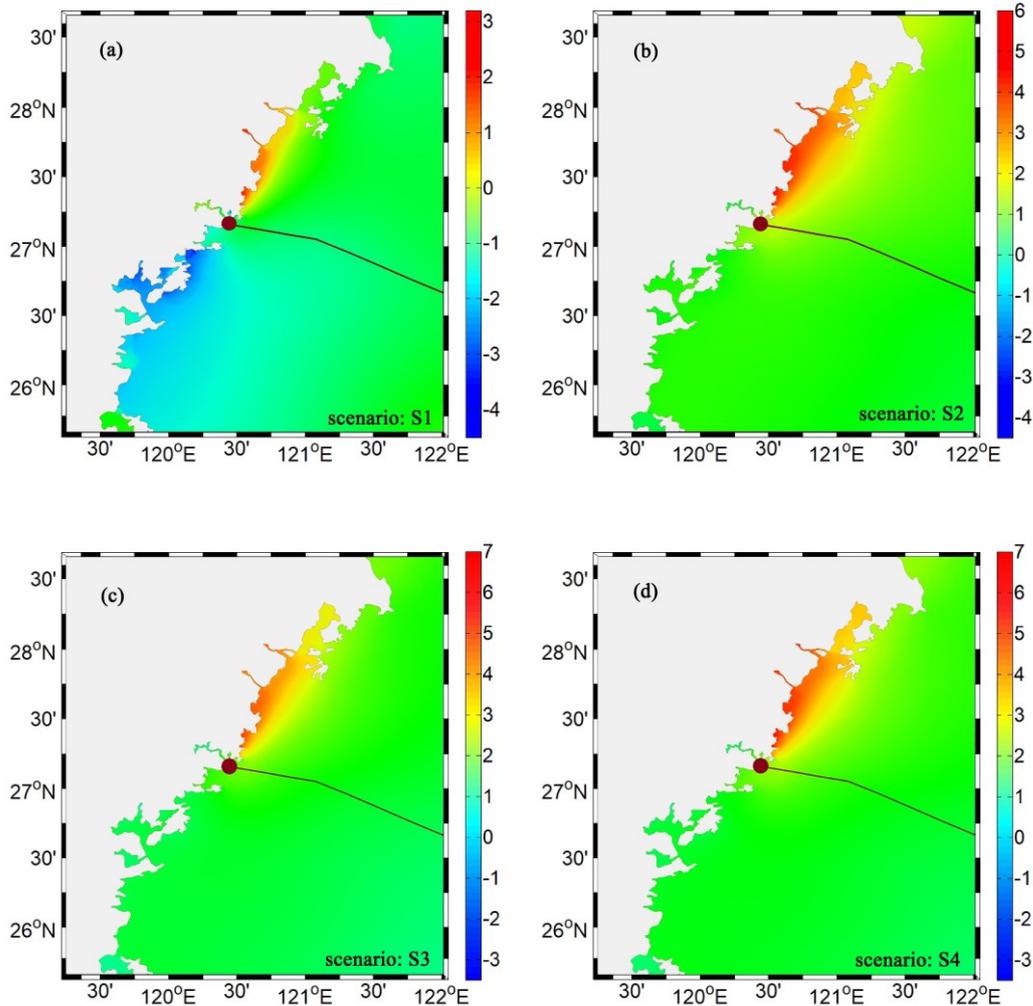

**Figure 5.** Water elevation (m) at landing moment. (a), (b), (c) and (d) are for secnario S1, S2, S3 and S4, respectively. The dot is the eye TC.

The distributions of water elevation at the landing moment for scenarios S1-S4 are presented as shown in figure 5. It indicates that the distributions of water elevation for the four scenarios are similar, but with significant difference in magnitude. Comparing S1 (see figures 5(a)) and S2 (see figure 5(b)), the landing moment has dramatic impact on the magnitude of water elevation. Specifically, the maximal water elevation for S2 is as high as about 6.0 m, while that for S1 is about 3.1 m. Significant difference of maximal water elevation along coastline between S1 and S2 can also be observed in figure 7(a). It implies that the potential storm tide hazard if Saomai landed during the period of astronomical high tide would be much more catastrophic than the actual situation of more than 480 deaths and 2.5 billion USD economic loss.

It is noteworthy that the difference of water elevation between S2 and S1 is not a linear superposition of storm surge on different astronomical tide levels. The tide-surge-wave coupling effect also play a role. Wuxi et al. (2018) concluded that the water elevation due to coupling effect experiences insignificant premonitory fluctuation, remarkable rise/drop and residual oscillating with a gentle decay of magnitude. To observe the distribution and magnitude of impact of coupling effect on water elevation, the difference of water elevation between S2 and S1 with the astronomical tide excluded is presented in figure 6. Difference of water elevation of −1 m maximal can be observed at area north of the TC track and in the bays. That means coupling effect can reduce the water elevation when astronomical tidal level increases. The main reasons are: 1) waves with larger height can survival in higher water depth resulting in less wave induced surge, 2) nonlinear tide-surge effect trends to reduce water elevation when astronomical tidal level increases.

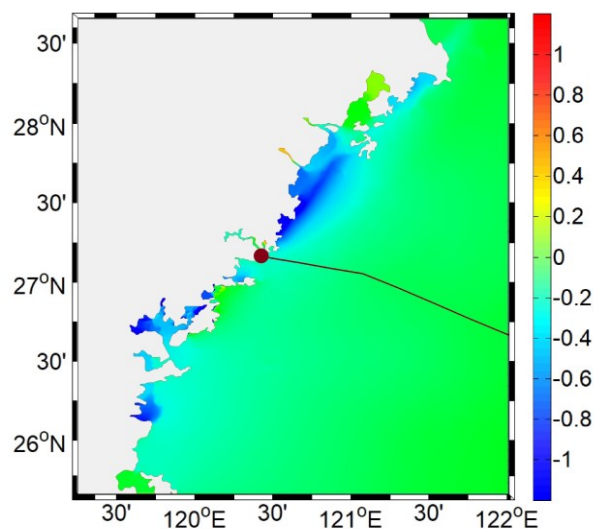

**Figure 6.** Difference of water elevation (m) between S2 and S1 with the astronomical tide excluded.

The maximal water elevations along the coastline north of the landing location during the whole process of TC passing by for S1-S4 are plotted in figure 7. From figure 7, the curves of water elevation along the coastline for S2-S4 can be divided into three segments with obvious different magnitudes of water elevation, i.e. the segment 0-25 km, the segment 25-225 km and the segment beyond 225 km. The increase of water elevation in the segment 0-25 km is mainly induced by the longshore current blocked by the convex of the coastline when the radius of maximum wind is close to the landing location; while that in the segment 25-225 km is due to the accumulation of water body blown by wind towards coast when the eye of TC is close to the landing location; for the coastline out of the radius of maximal wind, i.e. the segment beyond 225 km, water elevation is much smaller. Figure 7 shows that the average maximal water elevations along the coastline are 4.91 m, 5.25 m and 5.71 m for S2-S4, respectively. The comparison between S2 and S4 indicates that the non-stationary TCI and SLR effects increase water elevation by 0.80 m on average for the coastline concerned. More specifically, the water elevation due to SLR is a little bit smaller than 0.51 m, while that due to TCI is a little bit larger than 0.29 m. Because the MSL of S4 is 0.51 m larger than that of S2, and the tide-surge-wave coupling effects will reduce water elevation slightly when SLR, which is similar to the behavior of the coupling effect when astronomical tidal level increases (see figure 6). Comparing S3 and S4, it shows that the non-stationary TCI and SLR effects increase water elevation along the coastline by 0.46 m averagely, in which SLR effect is likely to share about 0.33 m. Accordingly, we can conclude that water elevation will be underestimated remarkably without considering the climate change impact, and both non-stationary TCI and SLR are important factors should be dealt with for long-term hazard assessment of storm tide.

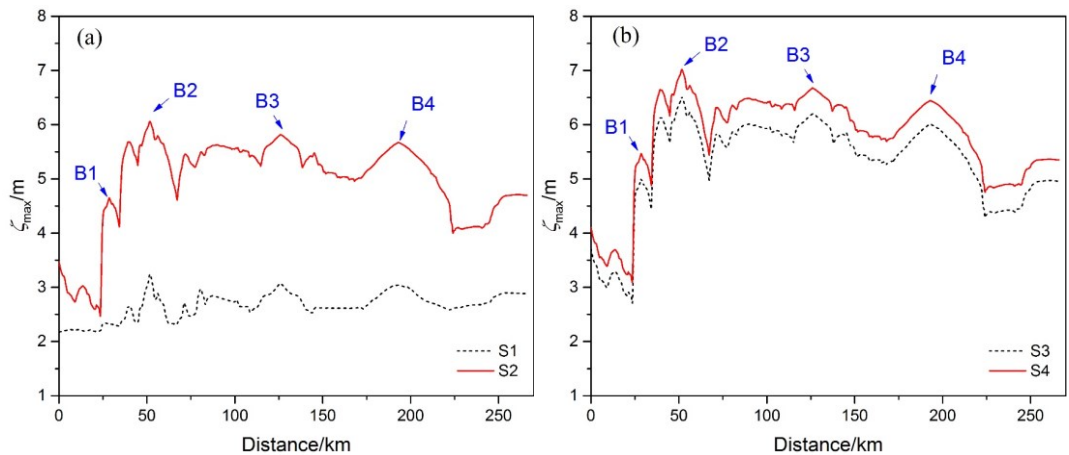

**Figure 7.** The maximal water elevation along the coastline during the whole process of TC landing. B1-B4 are the extremums of water elevation with their locations marked in figure 8. The horizontal axis represents the distance to the landing location of the TC eye.

Further, the locations of the extremums of water elevation curves in figure 7, i.e. B1-B4, are marked in figure 8. It demonstrates that high water elevation occurs in the bays or at the river estuaries, since water body is easy to accumulate in such geography. Specifically, B1 is the bay near the landing location of the TC eye; while B2-B4 are Aojiang river estuary, Feiyunjiang river estuary and Oujiang river estuary, respectively. The maximal water elevations at Aojiang river estuary (i.e. B2) are 6.06 m, 6.51 m and 7.02 m for S2-S4, respectively. While the maximal water elevations at Feiyunjiang river estuary (i.e. B3) are 5.82 m (S2), 6.20 m (S3) and 6.67 m (S4), respectively, and that at Oujiang river estuary (i.e. B4) are 5.67 m (S2), 6.00 m (S3) and 6.44 m (S4), respectively.

## 5.2 Storm tide inundation

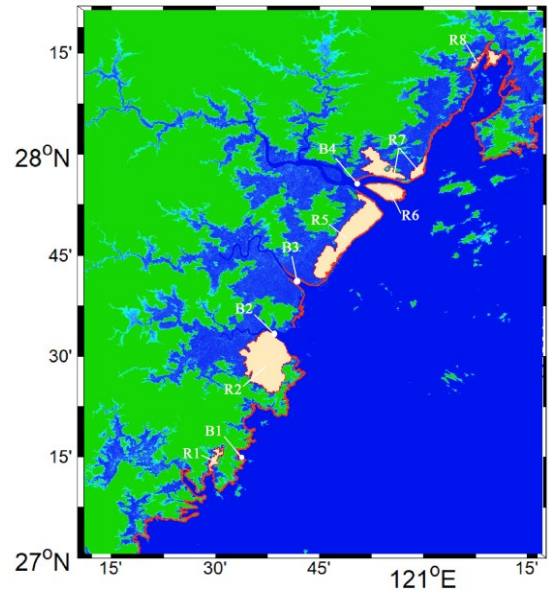

**Figure 8.** Overview of the potential inundation regions for S2. R1, R2 and R5-R8 are six regions with severe inundation hazard identified. B1-B4 are the locations of the extremums of water elevation in figure 7.

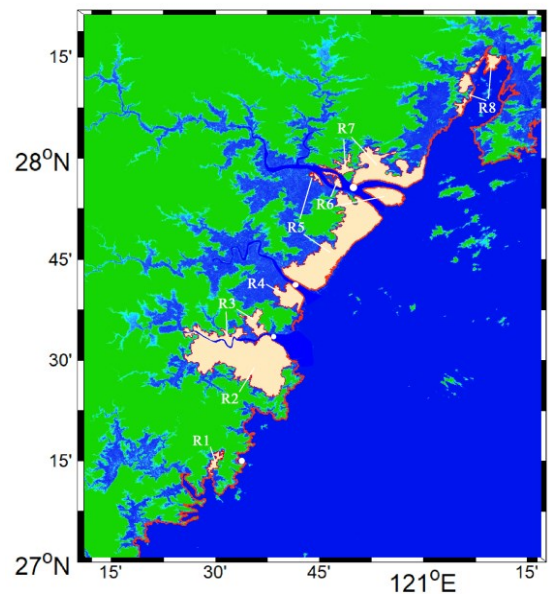

**Figure 9.** Overview of the potential inundation regions for S4. R1-R8 are eight regions with severe inundation hazard identified.

Relied on the GIS platform, the potential inundation regions with area larger than 10 km$^2$ for S2 and S4 are identified as shown in figures 8 and 9, respectively. From figure 8, the south coast of Aojiang river (R2), the south coast of Oujiang river (R5), the alluvial island at Oujiang river estuary (R6) and the north coast of Oujiang river (R7) are the largest four regions inundated. And low-lying regions R1 and R8 are also under the threat of inundation. As for S4 with the non-stationary TCI and SLR effects considered, the aforementioned six regions enlarge significantly. In addition, the north coast of Aojiang river (R3), the south and north coasts of Feiyunjiang river (R4 and R5) expose to the storm tide as well. The areas of the aforementioned inundation regions are presented in table 2, and the lengths of the coastlines where water body intruded are also presented along with. From table 2, about 533 km$^2$ is under the threat of storm tide inundation for S2. As for S4, the potential inundation will increase by 50% to about 798 km$^2$. That is to say, the storm tide inundation area will be underestimated remarkably as well without considering the non-stationary TCI and SLR effects.

Table 2. The area and length of the coastlines where water body intruded for the inundation regions.

|  |  | R1 | R2 | R3 | R4 | R5 | R6 | R7 | R8 |
|---|---|---|---|---|---|---|---|---|---|
| S2 | Area (km$^2$) | 10.2 | 144.7 | — | — | 122.1 | 30.5 | 54.6 | 17.1 |
| S2 | Length (km) | 1.6 | 13.9 | — | — | 32.8 | 25.3 | 15.5 | 17.8 |
| S4 | Area (km$^2$) | 10.2 | 273.9 | 47.0 | 35.3 | 235.0 | 37.6 | 117.8 | 41.5 |
| S4 | Length (km) | 1.9 | 13.9 | 3.3 | 10.5 | 53.9 | 32.6 | 25.8 | 30.0 |

Table 2 shows that the estuaries of Aojiang river (B2), Feiyunjiang river (B3) and Oujiang river (B4) are the hardest hit regions. The remotely sensed maps of the three hardest hit regions are presented in figure 10. The coastlines where water intruded are sketched by white solid lines, the farthest boundary of water body intruded for scenarios S2 and S4 are marked by red and blue lines, respectively. Figure 10(a) shows that the Cangnan city located at the south coast of Aojiang river is suffering from storm tide inundation hazard for S2. As for S4, the whole downtown of Cangnan city and almost half of Aojiang town located at the north coast of Aojiang river are exposed to the storm tide inundation. At the estuary of Feiyunjiang river (see figure 10(b)), most of the inundation areas are farmland. Figure 10(c) shows that most of the Longwan and Dongtou districts of Wenzhou city are of high risks of storm tide inundation. To provide reference for the local vulnerability assessments, the inundation durations for spots with different height over MSL for S2-S4 are also provided in figure 11. It shows that the inundation duration of S4 ranks the first followed by S3 and S2, which is consistent with their water elevations. Form figure 11, the inundation duration is not spatial homogeneous, and the estuary of Feiyunjiang river experiences longer period than the other two estuaries.

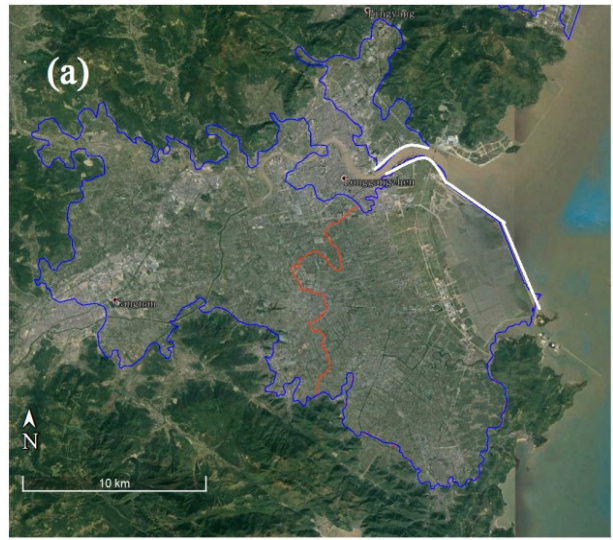

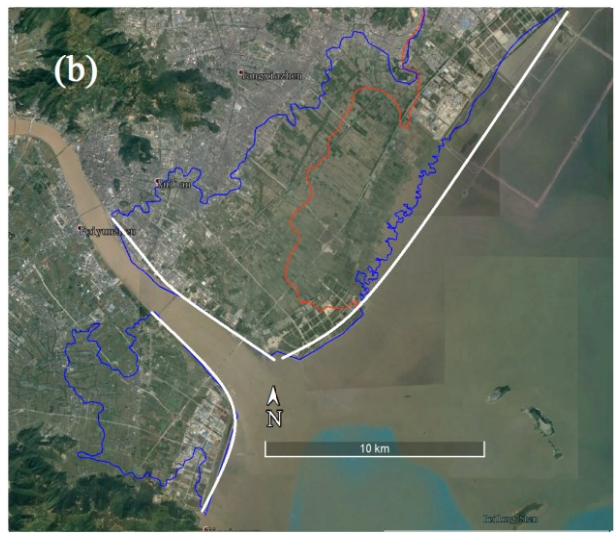

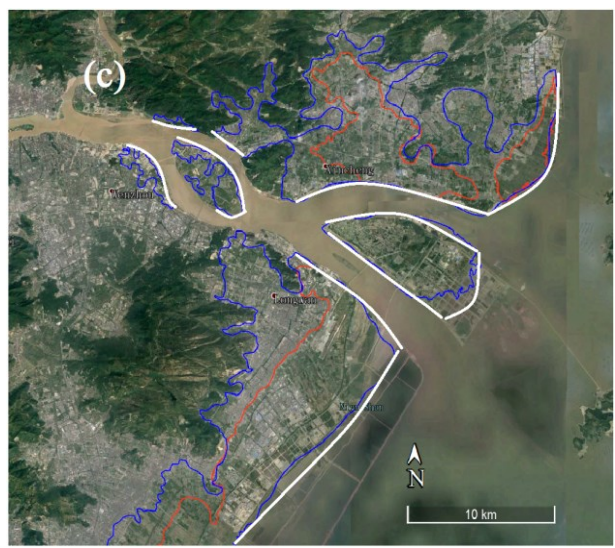

**Figure 10.** Remote sensed maps of hardest hit regions of storm tide inundation.
(a), (b) and (c) are for the estuaries of Aojiang river, Feiyunjiang river and Oujiang

river, respectively. The white line indicates the coastline where water body intruded. The red and blue lines correspond to the inundation lines for S2 and S4, respectively.

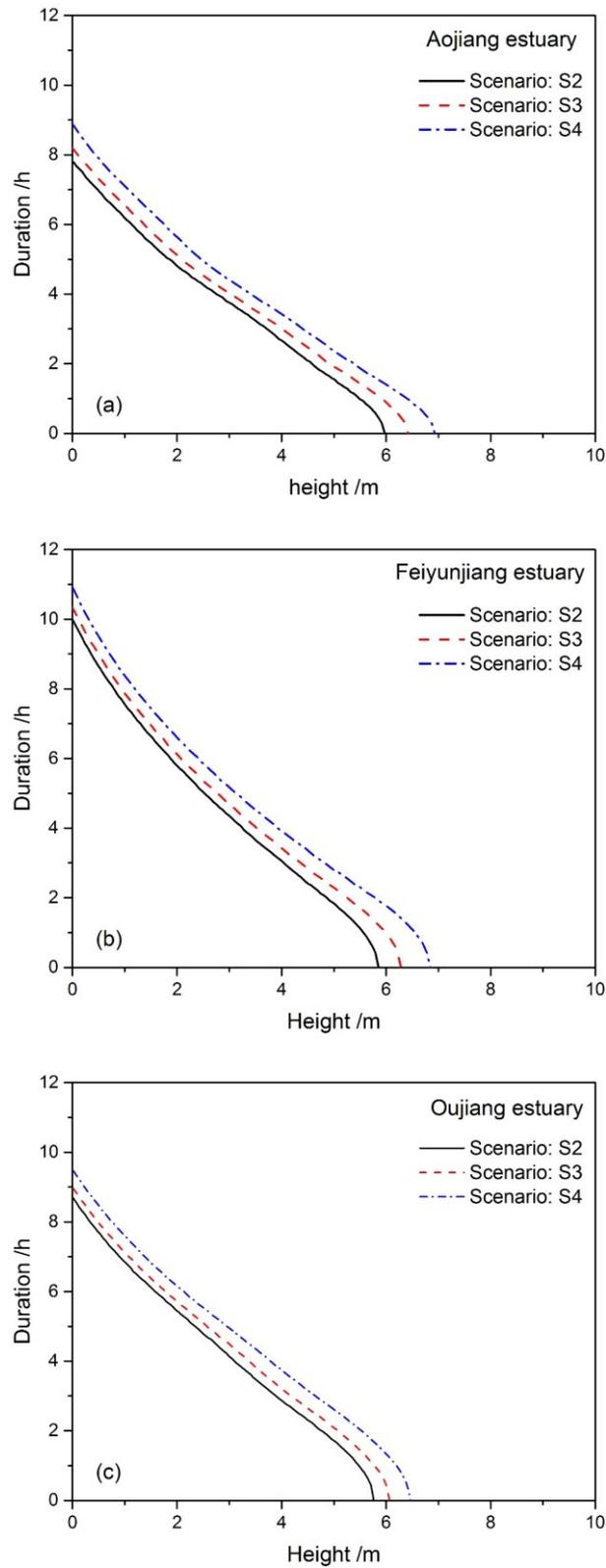

**Figure 11.** Inundation duration for the hardest hit regions of storm tide inundation. (a), (b) and (c) are for the estuaries of Aojiang river, Feiyunjiang river and

Oujiang river, respectively. The horizontal axis represents the height over the MSL. The solid line, dash line and dot dash line are for scenarios S2, S3 and S4, respectively.

## 6. Conclusions

Storm tide is the deadliest marine hazard, which has claimed large numbers of human lives and shocking economic losses. In the future, the potential storm tide inundation is by no means optimistic due to TCI and SLR. The storm tide inundation along Southeast China coast, one of the storm surge prone areas in China, is focused on in this study. Based on the statistical analysis of the historical data, the non-stationary TCI and SLR at the study area are estimated. Using the surge-tide-wave coupled model ADCIRC+SWAN, the evolution of water elevation is examined. Relied on the GIS platform, the hazard assessment of storm tide inundation and non-stationary TCI and SLR effects on that are presented and discussed.

The results demonstrate that the bays and estuaries in this coastal area tend to experience more dramatic rising process of water elevation. With effects of TCI and SLR neglected, the maximal water elevation around the estuary of Aojiang river by typhoon wind of 100-year recurrence period can be 6.06 m, while that around the Feiyunjiang and Oujiang river estuaries can be 5.82 m and 5.67 m, respectively. Under the same circumstances, roughly 533 $km^2$ of coast area is under the threat of storm tide inundation. Certainly, non-stationary TCI and SLR have remarkably impacts on the storm tide elevation and subsequent inundation. The water elevations caused by typhoon wind of 50-year recurrence period and 100-year recurrence period considering TCI and SLR are provided. The maximal water elevations caused by typhoon wind of 100-year recurrence period considering TCI and SLR can be as high as 6.51 m (Aojiang river estuary), 6.20 m (Feiyunjiang river estuary) and 6.00 m (Oujiang river estuary). The corresponding potential inundated area could expand by 50% to about 798 $km^2$. The remotely sensed maps for the most heavily hit regions, i.e. the estuaries of Aojiang, Feiyunjiang and Oujiang rivers are provided as well, which demonstrate that a few number of downtowns such as Cangnan city and Wenzhou city are exposed to the threat of storm tide inundation.

It is worth noting that the factors owing to climate change can be no longer neglected in the future risk assessment of storm tide disasters. Apart from the issues concerned in this study, relevant social and economic data are expected to be integrated in the present procedures for vulnerability assessment. Additionally, the robustness of the local system for immediate warning, efficient evacuating and in situ conservation need to be further evaluated in the future risk analysis as well.

# Acknowledgement

This work is supported by the Strategic Priority Research Programs (Category B) of the Chinese Academy of Sciences (XDB22040203) and the Opening Fund of the Hubei Key Laboratory of Naval Architecture & Ocean Engineering Hydrodynamics (201801). The authors also thank Guoqin Lyu for the assistance with the non-stationary estimation of extreme wind speed of TC.